\documentclass[twocolumn,showpacs,aps,prl,letterpaper]{revtex4}
\usepackage{graphicx}
\usepackage{dcolumn}
\usepackage{amsmath}
\usepackage{epsfig}
\long\def\inst#1{\par\nobreak\kern 4pt\nobreak
    {\itshape #1}\par\vskip 10pt plus 3pt minus 3pt}
\RequirePackage{xspace}
\usepackage{relsize}

\def\babar{\mbox{\slshape B\kern-0.1em{\smaller A}\kern-0.1em
    B\kern-0.1em{\smaller A\kern-0.2em R}}}
\def\Kbar    {\kern 0.18em\overline{\kern -0.18em K}{}\xspace}

\def\Kz      {\ensuremath{K^0}\xspace}
\def\Kzb     {\ensuremath{\Kbar^0}\xspace}
\def\KzKzb   {\ensuremath{\Kz {\kern -0.16em \Kzb}}\xspace}

\def\Ks     {\ensuremath{K_S}\xspace}
\def\Kl     {\ensuremath{K_L}\xspace}
\def\KsKs   {\ensuremath{\Ks {\kern -0.16em \Ks}}\xspace}
\def\KlKl   {\ensuremath{\Kl {\kern -0.16em \Kl}}\xspace}
\def\KsKl   {\ensuremath{\Ks {\kern -0.16em \Kl}}\xspace}
\def\KlKs   {\ensuremath{\Kl {\kern -0.16em \Ks}}\xspace}
\def\Dbar    {\kern 0.18em\overline{\kern -0.18em D}{}\xspace}
\def\Dz      {\ensuremath{D^0}\xspace}
\def\Dzb     {\ensuremath{\Dbar^0}\xspace}
\def\DzDzb   {\ensuremath{\Dz {\kern -0.16em \Dzb}}\xspace}
\def\Bbar    {\kern 0.18em\overline{\kern -0.18em B}{}\xspace}

\def\Bz      {\ensuremath{B^0}\xspace}
\def\Bzb     {\ensuremath{\Bbar^0}\xspace}
\def\BzBzb   {\ensuremath{\Bz {\kern -0.16em \Bzb}}\xspace}
\def\Bu      {\ensuremath{B^+}\xspace}
\def\Bub     {\ensuremath{B^-}\xspace}

\def\BpBm    {\ensuremath{\Bu {\kern -0.16em \Bub}}\xspace}

\newcommand{\optbar}[1]{\shortstack{{\tiny (\rule[.4ex]{1em}{.1mm})}
  \\ [-.7ex] $#1$}}
\def\BorBbar    {\kern 0.18em\optbar{\kern -0.18em B}{}\xspace}
\def\DorDbar    {\kern 0.18em\optbar{\kern -0.18em D}{}\xspace}
\def\KorKbar    {\kern 0.18em\optbar{\kern -0.18em K}{}\xspace}

\def\pep2{PEP-II}
\mathchardef\Upsilon="7107
\def\Y#1S{\ensuremath{\Upsilon{(#1S)}}\xspace}

\begin{document}

\title{\large \bfseries \boldmath Strong Phase and $\Dz-\Dzb$ mixing at
BES-III}
\author{Xiao-Dong Cheng$^{1,2}$}\email{chengxd@ihep.ac.cn}
\author{Kang-Lin He$^1$}\email{hekl@ihep.ac.cn}
\author{Hai-Bo Li$^1$}\email{lihb@ihep.ac.cn}
\author{Yi-Fang Wang$^1$}\email{yfwang@ihep.ac.cn}
\author{Mao-Zhi Yang$^1$}\email{yangmz@ihep.ac.cn}
\affiliation{$^1$Institute of High Energy Physics, P.O.Box 918,
Beijing  100049, China\\ $^2$Department of Physics, Henan Normal
University, XinXiang, Henan 453007, China}


\date{\today}


\begin{abstract}
Most recently, both BaBar and Belle experiments found evidences of
neutral $D$ mixing.  In this paper, we discuss the constraints on
the strong phase difference in $D^0 \rightarrow K\pi$ decay from
the measurements of the mixing parameters, $y^\prime$, $y_{CP}$
and $x$ at the $B$ factories. With $CP$ tag technique at
$\psi(3770)$ peak, the extraction of the strong phase difference
at BES-III are discussed. The sensitivity of the measurement of
the mixing parameter $y$ is estimated in BES-III experiment at
$\psi(3770)$ peak. Finally, we also make an estimate on the
measurements of the mixing rate $R_M$.

\end{abstract}


\maketitle


Due to the smallness of  $\Delta C=0$ amplitude in the Standard
Model (SM), $\Dz-\Dzb$ mixing offers a unique opportunity to probe
flavor-changing interactions which may be generated by new
physics. The recent measurements from BaBar and Belle experiments
indicate that the $\Dz-\Dzb$ mixing may exist~\cite{1,2}.  At the
$B$ factories, the decay time information can be used to extract
the neutral $D$ mixing parameters. At $t=0$ the only term in the
amplitude is the direct doubly-Cabibbo-suppressed (DCS) mode $D^0
\rightarrow K^+ \pi^-$, but for $t>0$ $\Dz-\Dzb$ mixing may
contribute through the sequence $D^0 \rightarrow \Dzb \rightarrow
K^+\pi^-$ , where the second stage is Cabibbo favored (CF).  The
interference of this term with the DCS contribution involves the
lifetime and mass differences of the neutral $D$ mass eigenstates,
as well as the final-state strong phase difference $\delta_{K\pi}$
between the CF and the DCS decay amplitudes. This interference
plays a key role in the measurement of the mixing parameters at
time-dependent measurements.


 With the assumption of $CPT$ invariance, the mass eigenstates of $\Dz -
\Dzb$ system are $|D_1\rangle = p |\Dz\rangle+q|\Dzb\rangle$ and
$|D_2\rangle = p |\Dz\rangle -q|\Dzb\rangle$ with eigenvalues
$\mu_1 = m_1 - \displaystyle\frac{i}{2} \Gamma_1$ and $\mu_2 = m_2
- \displaystyle\frac{i}{2} \Gamma_2$, respectively, where the
$m_1$ and $\Gamma_1$ ($m_2$ and $\Gamma_2$) are the mass and width
of $D_1$ ($D_2$). For the method of detecting $\Dz-\Dzb$ mixing
involving the $D^0 \rightarrow K\pi$ decay mentioned above, in
order to separate the DCS decay from the mixing signal, one must
study the time-dependent decay rate. The proper-time evolution of
the particle states $|\Dz_{\small\mbox{phys}}(t)\rangle$ and
$|\Dzb_{\small\mbox{phys}}(t)\rangle$ are given by
\begin{eqnarray}
|\Dz_{\small\mbox{phys}}(t) \rangle & = & g_+(t)  |\Dz \rangle
-\frac{q}{p}
g_-(t) |\Dzb \rangle, \nonumber \\
 |\Dzb_{\small\mbox{phys}}(t) \rangle & = & g_+(t) |\Dzb \rangle - \frac{p}{q} g_-(t) |\Dz \rangle,
\label{eq:d0_time}
\end{eqnarray}
where
\begin{eqnarray}
g_{\pm} = \frac{1}{2} (e^{-im_2 t-\frac{1}{2} \Gamma_2 t} \pm
e^{-im_1 t - \frac{1}{2}\Gamma_1 t} ),
 \label{eq:define}
\end{eqnarray}
with definitions
\begin{eqnarray}
m &\equiv& \frac{m_1 + m_2}{2}, \, \, \Delta m \equiv m_2 - m_1,
\nonumber \\
\Gamma &\equiv & \frac{\Gamma_1 + \Gamma_2}{2}, \, \Delta \Gamma
\equiv \Gamma_2 - \Gamma_1,
 \label{eq:define2}
\end{eqnarray}
Note  the sign of $\Delta m$ and $\Delta \Gamma$ is to be
determined by experiments.

 In practice, one define the following mixing parameters
\begin{eqnarray}
 x \equiv \frac{\Delta m}{\Gamma}, \, y \equiv \frac{\Delta
 \Gamma}{2\Gamma}.
\label{eq:define3}
\end{eqnarray}

The time-dependent decay amplitudes for
$\Dz_{\small\mbox{phys}}(t) \rightarrow K^+ \pi^-$ and
$\Dzb_{\small\mbox{phys}}(t) \rightarrow K^- \pi^+$ are described
as
\begin{eqnarray}
\langle K^+ \pi^- |{\cal H}|\Dz_{\small\mbox{phys}}(t) \rangle & =
& g_+(t) A_{K^+\pi^-} - \frac{q}{p} g_-(t) \overline{A}_{K^+\pi^-}
\nonumber \\
&=& \frac{q}{p} \overline{A}_{K^+\pi^-} [\lambda g_+(t) - g_-(t)],
 \label{eq:time:amplitude_D}
\end{eqnarray}
\begin{eqnarray}
\langle K^- \pi^+ |{\cal H}|\Dzb_{\small\mbox{phys}}(t) \rangle &
= & g_+(t) \overline{A}_{K^-\pi^+} - \frac{p}{q} g_-(t)
A_{K^-\pi^+}
\nonumber \\
&=& \frac{p}{q} A_{K^-\pi^+} [\overline{\lambda}g_+(t) - g_-(t)],
 \label{eq:time:amplitude_Dbar}
\end{eqnarray}
where $A_{K^+ \pi^-} = \langle K^+ \pi^- |{\cal H}|D^0 \rangle$,
$\overline{A}_{K^+ \pi^-} = \langle K^+ \pi^- |{\cal H}|\Dzb
\rangle$, $A_{K^- \pi^+} = \langle K^- \pi^+ |{\cal H}|D^0
\rangle$, and $\overline{A}_{K^- \pi^+} = \langle K^- \pi^+ |{\cal
H}|\Dzb \rangle$. Here, $\lambda$ and $\overline{\lambda}$ are
defined as:
\begin{eqnarray}
 \lambda \equiv \frac{p}{q}
 \frac{A_{K^+\pi^-}}{\overline{A}_{K^+\pi^-}},
 \label{eq:lambda}
\end{eqnarray}
\begin{eqnarray}
 \overline{\lambda} \equiv \frac{q}{p}
 \frac{\overline{A}_{K^-\pi^+}}{A_{K^-\pi^+}}.
 \label{eq:lambdabar}
\end{eqnarray}

From Eqs.~(\ref{eq:time:amplitude_D}) and
(\ref{eq:time:amplitude_Dbar}),  one can derive the general
expression for the time-dependent decay rate, in agreement
with~\cite{PDG2006,nir_2007}:
\begin{eqnarray}
\frac{d\Gamma(\Dz_{\small\mbox{phys}}(t) \rightarrow K^+
\pi^-)}{dt {\cal N}} &=&
|\overline{A}_{K^+\pi^-}|^2\left|\frac{q}{p}\right|^2 e^{-\Gamma
t}\times \nonumber \\
&& [(|\lambda|^2+1)\mbox{cosh}(y\Gamma t) + \nonumber \\
&& (|\lambda|^2 -1) \mbox{cos}(x\Gamma t) + \nonumber \\
&& 2 {\cal R}e(\lambda)\mbox{sinh}(y\Gamma t) + \nonumber \\
&& 2 {\cal I}m(\lambda)\mbox{sin}(x\Gamma t)]
 \label{eq:decay_rate_D}
\end{eqnarray}
\begin{eqnarray}
\frac{d\Gamma(\Dzb_{\small\mbox{phys}}(t) \rightarrow K^-
\pi^+)}{dt {\cal N}} &=&
|A_{K^-\pi^+}|^2\left|\frac{p}{q}\right|^2 e^{-\Gamma
t}\times \nonumber \\
&& [(|\overline{\lambda}|^2+1)\mbox{cosh}(y\Gamma t) + \nonumber \\
&& (|\overline{\lambda}|^2 -1) \mbox{cos}(x\Gamma t) + \nonumber \\
&& 2 {\cal R}e(\overline{\lambda})\mbox{sinh}(y\Gamma t) + \nonumber \\
&& 2 {\cal I}m(\overline{\lambda})\mbox{sin}(x\Gamma t)]
 \label{eq:decay_rate_Dbar}
\end{eqnarray}
where ${\cal N}$ is a common normalization factor. In order to
simplify the above formula, we make the following definition:
\begin{eqnarray}
\frac{q}{p} \equiv (1+A_M) e^{-i\beta},
 \label{eq:qp_par}
\end{eqnarray}
where $\beta$ is the weak phase in mixing and $A_M$ is a
real-valued parameter which indicates the magnitude of $CP$
violation in the mixing.  For $f = K^- \pi^+$ final state, we
define
\begin{eqnarray}
\frac{A_{K^+\pi^-}}{\overline{A}_{K^+\pi^-}} \equiv -
\sqrt{r^\prime} e^{-i \alpha^\prime}; \, \,
\frac{\overline{A}_{K^-\pi^+}}{A_{K^-\pi^+}} \equiv - \sqrt{r}
e^{-i \alpha},
 \label{eq:double_par}
\end{eqnarray}
where $r^\prime$ and $\alpha^\prime$ ($r$ and $\alpha$) are the
ratio and relative phase of the DCS decay rate and the CF decay
rate. Then, $\lambda$ and $\overline{\lambda}$ can be
parameterized as
\begin{eqnarray}
\lambda = -\sqrt{r^\prime} \frac{1}{1+A_M} e^{-i(\alpha^\prime
-\beta)} \, ,
 \label{eq:lambda_para}
\end{eqnarray}
\begin{eqnarray}
\overline{\lambda} = -\sqrt{r}(1+A_M) e^{-i(\alpha+\beta)}.
 \label{eq:lambda_para2}
\end{eqnarray}
In order to demonstrate the $CP$ violation in decay, we define
$\displaystyle \sqrt{r^\prime} \equiv \sqrt{R_D}(1+A_D)$ and
$\displaystyle \sqrt{r} \equiv \sqrt{R_D} \frac{1}{1+A_D}$. Thus,
Eqs. (\ref{eq:lambda_para}) and (\ref{eq:lambda_para2}) can be
expressed as
\begin{eqnarray}
\lambda = -\sqrt{R_D} \frac{1+A_D}{1+A_M} e^{-i(\delta -\phi)}\, ,
 \label{eq:lambda_para_s}
\end{eqnarray}
\begin{eqnarray}
\overline{\lambda} = -\sqrt{R_D}\frac{1+A_M}{1+A_D}
e^{-i(\delta+\phi)}\, ,
 \label{eq:lambda_para2_s}
\end{eqnarray}
where $\displaystyle\delta = \frac{\alpha^\prime +\alpha}{2}$ is
the averaged phase difference between DCS and CF processes, and
$\displaystyle\phi = \frac{ \alpha-\alpha^\prime}{2}+\beta$.

We can characterize the $CP$ violation in the mixing amplitude,
the decay amplitude, and the interference between amplitudes with
and without mixing, by real-valued parameters $A_M$, $A_D$, and
$\phi$ as in Ref~\cite{nir_1999,li_2006}. In the limit of $CP$
conservation, $A_M$, $A_D$ and $\phi$ are all zero.  $A_M = 0$
means  no $CP$ violation in mixing, namely, $|q/p|=1$; $A_D=0$
means no $CP$ violation in decay, for this case, $r = r^\prime =
R_D=
|\overline{A}_{K^-\pi^+}/A_{K^-\pi^+}|^2=|A_{K^+\pi^-}/\overline{A}_{K^+\pi^-}|^2$;
$\phi =0 $ means no $CP$ violation in the interference between
decay and mixing.

In experimental searches, one can define CF decay as right-sign
(RS) and DCS decay or via mixing followed by a CF decay as
wrong-sign (WS). Here, we define the ratio of WS to RS decays as
for $D^0$:
\begin{eqnarray}
R(t) = \frac{d\Gamma(\Dz_{\small\mbox{phys}}(t) \rightarrow K^+
\pi^-)}{dt {\cal N} \times e^{-\Gamma |t|} \times
2|\overline{A}_{K^+\pi^-}|^2} ,
 \label{eq:rt_d}
\end{eqnarray}
and for $\Dzb$:
\begin{eqnarray}
\overline{R}(t) =\frac{d\Gamma(\Dzb_{\small\mbox{phys}}(t)
\rightarrow K^- \pi^+)}{dt {\cal N}\times e^{-\Gamma |t|} \times
2|A_{K^-\pi^+}|^2} ,
 \label{eq:rt_dbar}
\end{eqnarray}

 Taking into account that $|\lambda|$,
 $|\overline{\lambda}| \ll 1$ and $x$, $y \ll 1$,
 keeping terms up to order $x^2$, $y^2$ and $R_D$ in the
expressions, neglecting $CP$ violation in  mixing,  decay and the
interference between decay with and without mixing ($A_M=0$,
$A_D=0$, and $\phi=0$), expanding the time-dependent for $x t$, $y
t \lesssim \Gamma^{-1}$, combing Eqs. (\ref{eq:decay_rate_D}) and
(\ref{eq:decay_rate_Dbar}), we can write Eqs. (\ref{eq:rt_d}) and
(\ref{eq:rt_dbar}) as
\begin{eqnarray}
R(t) = \overline{R}(t) = R_D + \sqrt{R_D} y^\prime \Gamma t +
\frac{x^{\prime^2} + y^{\prime^2}}{4} (\Gamma t)^2,
\label{eq:rt_exp}
\end{eqnarray}
where
\begin{eqnarray}
x^\prime &=& x \mbox{cos}\delta + y \mbox{sin} \delta,
\\
 y^\prime &=& -x \mbox{sin} \delta + y \mbox{cos} \delta.
\label{eq:x_y_phase}
\end{eqnarray}

In the limit of SU(3) symmetry, $A_{K^+ \pi^-}$ and
$\overline{A}_{K^+ \pi^-}$ ($A_{K^- \pi^+}$ and $\overline{A}_{K^-
\pi^+}$) are simply related by CKM factors, $A_{K^+ \pi^-}
=(V_{cd}V^*_{us}/V_{cs}V^*_{ud})\overline{A}_{K^+
\pi^-}$~\cite{grossman_2001}. In particular, $A_{K^+ \pi^-}$ and
$\overline{A}_{K^+ \pi^-}$ have the same strong phase, leading to
$\alpha^\prime = \alpha =0$ in Eq. (\ref{eq:double_par}).  But the
SU(3) symmetry is broken according to the recent precise
measurements from the $B$ factories, the ratio~\cite{nir_1999}:
\begin{eqnarray}
{\cal R} = \frac{{\cal BR}(D^0 \rightarrow K^+\pi^-)}{{\cal BR}
(\Dzb \rightarrow K^+\pi^-)} \left| \frac{V_{ud}V^*_{cs}}{V_{us}
V^*_{cd}} \right|^2,
 \label{eq:rt_su3}
\end{eqnarray}
is unity in the SU(3) symmetry limit. But, the world average for
this ratio is
\begin{eqnarray}
{\cal R}_{exp} = 1.21\pm 0.03,
 \label{eq:rt_su3}
\end{eqnarray}
computed from the individual measurements using the standard
method of Ref.~\cite{PDG2006}.  Since the SU(3) is broken in $D
\rightarrow K\pi$ decays at the level of 20\%, in which case the
strong phase $\delta$ should be non-zero. Recently, a
time-dependent analysis in $D \rightarrow K\pi$ has been performed
based on 384 fb$^{-1}$ luminosity at $\Upsilon(4S)$~\cite{1}. By
assuming $CP$ conservation, they obtained the following neutral
$D$ mixing results
\begin{eqnarray}
R_D &=& (3.03 \pm 0.16 \pm 0.10)\times 10^{-3} , \nonumber \\
x^{\prime^2}& =& (-0.22 \pm 0.30 \pm 0.21) \times 10^{-3},
\nonumber
\\
y^\prime &=& (9.7 \pm 4.4 \pm 3.1)\times 10^{-3}.
 \label{eq:kp_babar}
\end{eqnarray}
\begin{table}[htbp]
  \centering
  \caption{ Experimental results used in the paper. Only one error is quoted,
  we have combined in quadrature statistical and systematic contributions. }
  \begin{tabular}{c|c|c|c} \hline\hline
  Parameter & BaBar ($\times 10^{-3}$)  & Belle($\times 10^{-3}$) &  Technique \\ \hline
  $x^{\prime^2}$ & -$0.22\pm 0.37$~\cite{1} &$0.18^{+0.21}_{-0.23}$~\cite{belle_kp_06} & $K\pi$ \\
  $y^{\prime}$ & $9.7\pm 5.4$~\cite{1} & $0.6^{+4.0}_{-3.9}$~\cite{belle_kp_06} & $K\pi$ \\
  $R_D$ & $3.03\pm 0.19$~\cite{1} & $3.64\pm 0.17$~\cite{belle_kp_06} & $K\pi$ \\
  $y_{CP}$ & - & $13.1\pm 4.1$~\cite{2} & $K^+K^-$, $\pi^+\pi^-$ \\
  $x$ & -& $8.0\pm3.4$~\cite{marko_belle_07} & $K_S \pi^+\pi^-$ \\
  $y$ & -& $3.3\pm 2.8$~\cite{marko_belle_07} & $K_S \pi^+\pi^-$ \\
  \hline \hline
 \end{tabular}
  \label{tab:experiments}
\end{table}

The result is inconsistent with the no-mixing hypothesis with a
significance of 3.9 standard deviations.  The results from BaBar
and Belle are in agreement within 2 standard deviation on the
exact analysis of $y^\prime$ measurement by using $D \rightarrow
K\pi$ as listed in Table~\ref{tab:experiments}. As indicated in
Eq.~(\ref{eq:rt_su3}), the strong phase $\delta$ should be
non-zero due to the SU(3) violation.  One has to know the strong
phase difference exactly in order to extract the direct mixing
parameters, $x$ and $y$ as defined in Eqs. (\ref{eq:define3}).
However, at the $B$ factory, it is hard to do that with a
model-independent way~\cite{grossman_2001,ian_2003}. In order to
extract the strong phase $\delta$ we need data near the
$D\overline{D}$ threshold to do a $CP$ tag as discussed in
Ref.~\cite{grossman_2001}. Here, we would like to figure out the
possible physics solution of the strong phase $\delta$ by using
the recent results from the $B$ factories with different decay
modes,  so that we can have an idea about the sensitivity to
measure the strong phase at the BES-III project.

In Ref~\cite{2}, Belle collaboration also reported the result of
$y_{CP}=\frac{\tau(D^0\rightarrow K^+\pi^-)}{\tau(D^0 \rightarrow
f_{CP})} -1$,  where $f_{CP} = K^+K^-$ and $\pi^+\pi^-$
\begin{eqnarray}
 y_{CP} = (13.1 \pm 3.2 \pm 2.5) \times 10^{-3}.
 \label{eq:kk_belle}
\end{eqnarray}
The result is about 3.2$\sigma$ significant deviation from zero
(non-mixing). In the limit of $CP$ symmetry, $y_{CP} =
y$~\cite{nir_2000,petrov_2005}. In the decay of $D^0 \rightarrow
K_S \pi^+\pi^-$, Belle experiment has done a Dalitz plot (DP)
analysis~\cite{marko_belle_07}, they obtained the direct mixing
parameters $x$ and $y$ as
\begin{eqnarray}
 x = (8.0 \pm 3.4) \times 10^{-3}, \,\, y = (3.3 \pm 2.8)\times
 10^{-3},
 \label{eq:kspp_belle}
\end{eqnarray}
where the error includes both statistic and systematic
uncertainties. Since the parameterizations  of the resonances on
the DP are model-dependent,  the results suffer from large
uncertainties from the DP model. In this analysis, they see a
significance of 2.4 standard deviations from non-mixing. Here, we
will use the value of $x$ measured in the DP analysis for further
discussion. As shown in Eq. (\ref{eq:x_y_phase}), once $y$,
$y^\prime$ and $x$ are known, it is straightforward to extract the
strong phase difference between DCS and CF decay in $D^0
\rightarrow K\pi$ decay. If taking the measured central values of
$x$, $y_{CP}(\approx y)$ , and $y^\prime$ as input parameters, we
found two-fold solutions for $\mbox{tan}\delta$ as below:
\begin{eqnarray}
 \mbox{tan}\delta = 0.35\pm 0.63, \,\, \mbox{or} \, \, -7.14 \pm 29.13,
 \label{eq:phase_solution}
\end{eqnarray}
which are corresponding to $(19 \pm 32)^0$ and $(-82^0 \pm 30)^0$,
respectively.

At $\psi(3770)$ peak, to extract the mixing parameter $y$, one can
make use of rates for exclusive $D^0\Dzb$ combination, where both
the $D^0$ final states are specified (known as double tags or DT),
as well as inclusive rates, where either the $D^0$ or $\Dzb$ is
identified and the other $D^0$ decays generically (known as single
tags or ST)~\cite{asner_2005}.  With the DT tag
technique~\cite{markiii_1,markiii_2}, one can fully consider the
quantum correlation in $C=-1$ and $C=+1$ $D^0\Dzb$ pairs produced
in the reaction $e^+e^- \rightarrow D^0 \Dzb(n\pi^0)$ and $e^+e^-
\rightarrow D^0 \Dzb \gamma
(n\pi^0)$~\cite{bigi_tau,bigi_sanda,asner_2005}, respectively.

For the ST, in the limit of $CP$ conservation, the rate of $D^0$
decays into a $CP$ eigenstate is given as~\cite{asner_2005}:
\begin{eqnarray}
 \Gamma_{f_\eta}\equiv\Gamma(D^0 \rightarrow f_{\eta}) = 2A_{f_{\eta}}^2
\left[1-\eta  y \right],
 \label{eq:st_cp}
\end{eqnarray}
where $f_{\eta}$ is a $CP$ eigenstate with eigenvalue $\eta = \pm
1$, and $A_{f_{\eta}}= |\langle f_\eta | {\cal H}|D^0 \rangle|$ is
the real-valued decay amplitude.

For the DT case, Gronau {\it et. al.}~\cite{grossman_2001} and
Xing~\cite{xing_1997} have considered time-integrated decays into
correlated pairs of states, including the effects of non-zero
final state phase difference. As discussed in
Ref.~\cite{grossman_2001}, the rate of ($D^0 \Dzb)^{C=-1}
\rightarrow (l^\pm X)(f_\eta)$ is described
as~\cite{grossman_2001}:
\begin{eqnarray}
\Gamma_{l;f_\eta}\equiv \Gamma[(l^\pm X)(f_{\eta})] &= & A_{l^\pm
X}^2A_{f_\eta}^2 (1+ y^2) \nonumber \\
&\approx&  A_{l^\pm X}^2 A_{f_\eta}^2, \label{eq:dt_cp}
\end{eqnarray}
where $A_{l^\pm X} = | \langle l^\pm X|{\cal H}|D^0\rangle|$ is
real-valued amplitude for semileptonic decays, here, we neglect
$y^2$ term since $y\ll1$.

For $C=-1$ initial $D^0\Dzb$ state,  $y$ can be expressed in term
of the ratios of DT rates and the double ratios of ST rates to DT
rates~\cite{asner_2005}:
\begin{eqnarray}
y = \frac{1}{4 } \left( \frac{\Gamma_{l; f_+}
\Gamma_{f_-}}{\Gamma_{l;f_-}\Gamma_{f_+}} -\frac{\Gamma_{l; f_-}
\Gamma_{f_+}}{\Gamma_{l;f_+}\Gamma_{f_-}} \right ).
\label{eq:dt_cp_y}
\end{eqnarray}
For a small $y$, its error, $\Delta (y)$, is approximately
$1/\sqrt{N_{l^\pm X}}$, where $N_{l^\pm X}$ is the total number of
$(l^\pm X)$ events tagged with $CP$-even and $CP$-odd eigenstates.
The number $N_{l^\pm X}$ of $CP$ tagged events is related to the
total number of $D^0 \Dzb$ pairs $N(D^0 \Dzb)$ through $N_{l^\pm
X} \approx N(D^0 \Dzb)[ {\cal BR}(D^0 \rightarrow l^\pm +X)\times
{\cal BR}(D^0 \rightarrow f_{\pm})\times \epsilon_{tag}] \approx
1.5\times 10^{-3} N(D^0 \Dzb)$, here we take the branching
ratio-times-efficiency factor (${\cal BR}(D^0 \rightarrow
f_{\pm})\times \epsilon_{tag}$) for tagging $CP$ eigenstates is
about 1.1\% (the total branching ratio into $CP$ eigenstates is
larger than about 5\%~\cite{PDG2006}). We find
\begin{eqnarray}
\Delta(y) = \frac{\pm 26}{\sqrt{N(D^0\Dzb)}} = \pm 0.003.
 \label{eq:dt_cp_dy}
\end{eqnarray}
If we take the central value of $y$ from the measurement of
$y_{CP}$ at Belle experiment~\cite{2},  thus, at BES-III
experiment~\cite{besiii}, with 20$fb^{-1}$ data at $\psi(3770)$
peak, the significance of the measurement of $y$ could be around
4.3 $\sigma$ deviation from zero.

\begin{figure}
 \epsfig{file=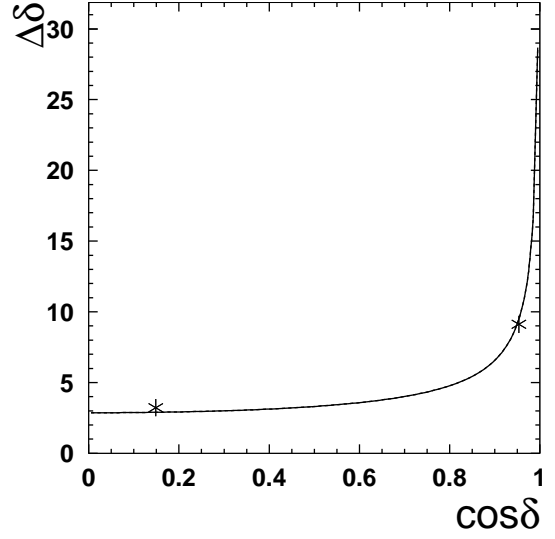,width=8cm,height=8cm}
\caption{Illustrative plot of the expected error ($\Delta \delta$)
of the strong phase with various central values of
$\mbox{cos}\delta$.  The expected error of $\mbox{cos}\delta$ is
0.04 by ssuming $20fb^{-1}$ data at $\psi(3770)$ peak at BES-III.
The two asterisks correspond to $\delta = 19^0$ and $-82^0$,
respectively.} \label{figure1}
\end{figure}

 We can also take advantage of the coherence of the $D^0$
mesons produced at the $\psi(3770)$ peak to extract the strong
phase difference $\delta$ between DCS and CF decay amplitudes that
appears in the time-dependent mixing measurement in
Eq.~(\ref{eq:rt_exp})~\cite{grossman_2001,asner_2005}. Because the
$CP$ properties of the final states produced in the decay of the
$\psi(3770)$ are anti-correlated~\cite{bigi_tau,bigi_sanda}, one
$D^0$ state decaying into a final state with definite $CP$
properties immediately identifies or tags the $CP$ properties of
the other side.  As discussed in Ref.~\cite{grossman_2001}, the
process of one $D^0$ decaying to $K^-\pi^+$, while the other $D^0$
decaying to a $CP$ eigenstate $f_{\eta}$  can be described as
\begin{eqnarray}
\Gamma_{K\pi;f_\eta}\equiv \Gamma[(K^- \pi^+)(f_{\eta})] &\approx
& A^2A^2_{f_{\eta}}|1+ \eta
\sqrt{R_D} e^{-i \delta} |^2  \nonumber \\
&\approx&  A^2A^2_{f_{\eta}}(1+2 \eta \sqrt{R_D}
\mbox{cos}\delta),\nonumber \\
 \label{eq:besiii_delta_rD}
\end{eqnarray}
where $A = |\langle K^- \pi^+ |{\cal H}| D^0 \rangle |$ and
$A_{f_{\eta}} = |\langle f_{\eta} |{\cal H}| D^0 \rangle |$ are
the real-valued decay amplitudes, and we have neglected the $y^2$
terms in Eq.~(\ref{eq:besiii_delta_rD}). In order to estimate the
total sample of events needed to perform a useful measurement of
$\delta$,  one defined~\cite{grossman_2001,ian_2003} an asymmetry
\begin{eqnarray}
{\cal A} \equiv \frac{\Gamma_{K\pi;f_+} - \Gamma_{K\pi;
f_-}}{\Gamma_{K\pi;f_+} +\Gamma_{K\pi; f_-}},
\label{eq:besiii_delta_rD_a}
\end{eqnarray}
where $\Gamma_{K\pi;f_\pm}$ is defined in
Eq.~(\ref{eq:besiii_delta_rD}), which is the rates for the
$\psi(3770) \rightarrow D^0 \Dzb$ configuration to decay into
flavor eigenstates and a $CP$-eigenstates $f_\pm$.
Eq.~(\ref{eq:besiii_delta_rD}) implies a small asymmetry, ${\cal
A} = 2 \sqrt{R_D} \mbox{cos} \delta$. For a small asymmetry, a
general result is that its error $\Delta {\cal A}$ is
approximately $1/\sqrt{N_{K^-\pi^+}}$, where $N_{K^-\pi^+}$ is the
total number of events tagged with $CP$-even and $CP$-odd
eigenstates. Thus one obtained
\begin{eqnarray}
\Delta (\mbox{cos} \delta) \approx \frac{1}{2\sqrt{R_D}
\sqrt{N_{K^-\pi^+}}}. \label{eq:besiii_delta_rD_est}
\end{eqnarray}
The expected number $N_{K^-\pi^+}$ of $CP$-tagged events can be
connected to the total number of $D^0 \Dzb$ pairs $N(D^0 \Dzb)$
through $N_{K^-\pi^+} \approx N(D^0 \Dzb){\cal BR}(D^0 \rightarrow
K^- \pi^+) \times {\cal BR}(D^0 \rightarrow f_{\pm})\times
\epsilon_{tag} \approx 4.2 \times 10^{-4} N(D^0
\Dzb)$~\cite{grossman_2001}, here, as in Ref~\cite{grossman_2001},
we take the branching ratio-times-efficiency factor ${\cal BR}(D^0
\rightarrow f_{\pm})\times \epsilon_{tag}=1.1\%$.  With the
measured $R_{D} = (3.03\pm 0.19)\times 10^{-3}$ and ${\cal BR}(D^0
\rightarrow K^- \pi^+)=3.8\%$~\cite{PDG2006}, one
found~\cite{grossman_2001}
\begin{eqnarray}
\Delta (\mbox{cos}\delta) \approx \frac{\pm 444}{\sqrt{N(D^0
\Dzb)}}. \label{eq:besiii_delta_rD_est_num}
\end{eqnarray}
At BESIII,  about $72 \times 10^6$ $D^0 \Dzb$ pairs can be
collected with 4 years' running. If considering both $K^- \pi^+$
and $K^+\pi^-$ final states,  we thus estimate that one may be
able to reach an accuracy of about 0.04 for cos$\delta$.
Figure~\ref{figure1} shows the expected error of the strong phase
$\delta$ with various central values of $\mbox{cos}\delta$. With
the expected $\Delta(\mbox{cos}\delta)= \pm 0.04$,  the
sensitivity of the strong phase varies  with the physical value of
$\mbox{cos}\delta$. For $\delta = 19^0$ and $-82^0$, the expected
error could be $\Delta(\delta) = \pm 8.7^0$ and $\pm 2.9^0$,
respectively.

By combing the measurements of $x$ in $D^0 \rightarrow K_S \pi\pi$
and $y_{CP}$ from Belle, one can obtain $R_M =(1.18\pm 0.6)\times
10^{-4}$. At the $\psi(3770)$ peak, $D^0\Dzb$ pair are produced in
a state that is quantum-mechanically
coherent~\cite{bigi_tau,bigi_sanda}. This enables simple new
method to measure $D^0$ mixing parameters in a way similar
proposed in Ref.~\cite{grossman_2001}. At BES-III, the measurement
of $R_M$ can be performed unambiguously with the following
reactions~\cite{bigi_tau}:
\begin{eqnarray}
&(i)&\,\, e^+ e^- \rightarrow \psi(3770) \rightarrow D^0 \Dzb
\rightarrow
(K^\pm \pi^\mp)( K^\pm \pi^\mp), \nonumber \\
&(ii)&\, \, e^+ e^- \rightarrow \psi(3770) \rightarrow D^0 \Dzb
\rightarrow
(K^- e^+ \nu )(K^- e^+ \nu),  \nonumber \\
&(iii)& \, \, e^+e^- \rightarrow D^- D^{*+} \rightarrow (K^+ \pi^-
\pi^-) ( \pi^+_{soft} [K^+ e^- \nu]). \nonumber \\
\label{eq:besiii_rm_corr_1}
\end{eqnarray}
Reaction $(i)$ in Eq.~(\ref{eq:besiii_rm_corr_1}) can be
normalized to $D^0\Dzb \rightarrow (K^-\pi^+)(K^+\pi^-)$, the
following time-integrated ratio is obtained by neglecting $CP$
violation:
\begin{eqnarray}
\frac{N[(K^- \pi^+ )(K^-\pi^+)]}{N[(K^-\pi^+)(K^+\pi^-)]} \approx
\frac{x^2+y^2}{2} = R_M. \label{eq:besiii_rm_measure_hadron}
\end{eqnarray}
For the case of semileptonic decay, as $(ii)$ in
Eq.~(\ref{eq:besiii_rm_corr_1}),  we have
\begin{eqnarray}
\frac{N(l^{\pm} l^\pm)}{N(l^\pm l^\mp)} = \frac{x^2+y^2}{2} = R_M,
\label{eq:besiii_rm_measure}
\end{eqnarray}

The observation of reaction $(i)$  would be definite evidence for
the existence of $D^0 -\Dzb$ mixing since the final state $(K^\pm
\pi^\mp)( K^\pm \pi^\mp)$ can not be produced from DCS decay due
to quantum statistics~\cite{bigi_sanda,bigi_tau}. In particular,
the initial $D^0\Dzb$ pair is in an odd eigenstate of $C$ which
will preclude, in the absence of mixing between the $D^0$ and
$\Dzb$ over time, the formation of the symmetric state required by
Bose statistics if the decays are to be the same final state. This
final state is also very appealing experimentally, because it
involves a two-body decay of both charm mesons, with energetic
charged particles in the final state that form an overconstrained
system. Particle identification is crucial in this measurement
because if both the kaon and pion are misidentified in one of the
two $D$-meson decays in the event, it becomes impossible to
discern whether mixing has occurred. At BESIII, where the data
sample is expected to be 20 fb$^{-1}$ integrated luminosity at
$\psi(3770)$ peak,  the limit will be $10^{-4}$ at 95\% C.L. for
$R_M$, but only if the particle identification capabilities are
adequate.

Reactions $(ii)$ and $(iii)$ offer unambiguous evidence for the
mixing because the mixing is searched for in the semileptonic
decays for which there are no DCS decays. Of course since the
time-evolution is not measured, observation of Reactions $(ii)$
and $(iii)$ actually would indicate the violation of the selection
rule relating the change in charm to the change in leptonic charge
which holds true in the standard model~\cite{bigi_tau}.

In Table~\ref{tab:mixing_r_m}, the sensitivity for $R_M$
measurements in different decay modes are estimated with 4 years'
run at BEPCII.
\begin{table}[htbp]
  \centering
\caption{The sensitivity for $R_M$ measurements at BES-III with
different decay modes with 4 years' run at BESPCII}
  \begin{tabular}{c|c|c} \hline \hline
 \multicolumn{3}{c}{$D^0\Dzb$ Mixing} \\ \hline

 Reaction & Events  & Sensitivity  \\
         & RS($\times 10^{4}$) &   $R_M$($\times 10^{-4}$)               \\ \hline
 $\psi(3770) \rightarrow (K^-\pi^+)(K^-\pi^+)$ & 10.4  & $ 1.0$\\ \hline
 $\psi(3770) \rightarrow (K^- e^+ \nu)(K^- e^+ \nu )$ & 8.9 &   \\
 $\psi(3770) \rightarrow (K^- e^+ \nu)(K^- \mu^+ \nu )$ & 8.1  & $ 3.7$\\
 $\psi(3770) \rightarrow (K^- \mu^+ \nu)(K^- \mu^+ \nu )$ & 7.3  &  \\
\hline \hline
\end{tabular}
\label{tab:mixing_r_m}
\end{table}

In the limit of $CP$ conservation, by combing the measurements of
$x$ in $D^0 \rightarrow K_S \pi\pi$ and $y_{CP}$ from Belle,  one
can obtain $R_M =(1.18\pm 0.6)\times 10^{-4}$.  With 20fb$^{-1}$
data at BES-III, about 12 events for the precess $D^0\Dzb
\rightarrow (K^\pm \pi^\mp) (K^\pm \pi^\mp)$ can be produced. One
can observe 3.0 events after considering the selection efficiency
at BESIII, which could be about 25\% for the four charged
particles. The background contamination due to double particle
misidentification is about 0.6 event with 20$fb^{-1}$ data at
BES-III~\cite{kanglin_07}. Table~\ref{tab:possion} lists the
expected mixing signal for
$N_{sig}=N(K^\pm\pi^\mp)(K^\pm\pi^\mp)$, background $N_{bkg}$ ,
and the Poisson probability $P(n)$, where $n$ is the possible
number of observed events in experiment. In
Table~\ref{tab:possion}, we assume the $R_M =1.18\times 10^{-4}$,
the expected number of mixing signal events are estimated with
10fb$^{-1}$ and 20fb$^{-1}$, respectively.
\begin{table}[htbp]
  \centering
\caption{The expected mixing signal for
$N_{sig}=N(K^\pm\pi^\mp)(K^\pm\pi^\mp)$, background $N_{bkg}$ ,
and the Poisson probability $P(n)$ in 10 fb$^{-1}$ and 20
fb$^{-1}$ at BES-III at $\psi(3770)$ peak, respectively. Here, we
take the mixing rate $R_M=1.18\times 10^{-4}$. }
  \begin{tabular}{c|c|c} \hline \hline
   & 10 fb$^{-1}$ ($\psi(3770)$)  &   20 fb$^{-1}$ ($\psi(3770)$) \\
   & 36 million $D^0\Dzb$ &72 million
$D^0\Dzb$ \\ \hline
  $N_{sig}$  & 1.5 & 3.0      \\
  $N_{bkg}$  & 0.3 & 0.6      \\ \hline
  $P(n=0)$  & 15.7\% & 2.5\%  \\
 $P(n=1)$   & 29.1\% & 9.1\%  \\
 $P(n=2)$  & 26.9\% & 16.9\%  \\
 $P(n=3)$   & 16.6\% & 20.9\%  \\
 $P(n=4)$  & 7.7\% & 19.3\%  \\
 $P(n=5)$   & 2.8\% & 14.3\%  \\
  $P(n=6)$  & 0.9\% & 8.8\%  \\
 $P(n=7)$   & 0.2\% & 4.7\%  \\
  $P(n=8)$  & 0.1\% & 2.2\%  \\
 $P(n=9)$   & 0.01\% & 0.9\%  \\ \hline \hline
\end{tabular}
\label{tab:possion}
\end{table}

In conclusion,  we discuss the constraints on the strong phase
difference in $D^0 \rightarrow K\pi$ decay according to the most
recent measurements of $y^\prime$, $y_{CP}$ and $x$ from $B$
factories. We estimate the sensitivity of the measurement of
mixing parameter $y$ at $\psi(3770)$ peak in BES-III experiment.
With 20 fb$^{-1}$ data, the uncertainty $\Delta(y)$ could be
0.003. Thus, assuming $y$ at a percent level, we can make a
measurement of $y$ at a significance of 4.3$\sigma$ deviation from
zero. The sensitivity of the strong phase difference at BES-III
are obtained by using data near the $D\overline{D}$ threshold with
$CP$ tag technique at BES-III experiment. Finally, we estimated
the sensitivity of the measurements of the mixing rate $R_M$, and
find that BES-III experiment may not be able to make a significant
measurement of $R_M$ with current luminosity by using coherent
$D\overline{D}$ state at $\psi(3770)$ peak.

One of the authors (H.~B.~Li) would like to thank David Asner and Zhi-Zhong Xing
for stimulating discussion,  Chang-Zheng Yuan for useful
discussion on the statistics used in this paper, and also thank
Stephen L. Olsen and Yang-Heng Zheng for commenting on this
manuscript. We thank BES-III collaboration for providing us many
numerical results based on GEANT4 simulation. This work is
supported in part by the National Natural Science Foundation of
China under contracts Nos. 10205017, 10575108,10521003, and the
Knowledge Innovation Project of CAS under contract Nos. U-612 and
U-530 (IHEP).




\end{document}